\documentclass[preprint2]{aastex}

\def\dsm{$M_\odot$}
\def\dsr{$R_\odot$}
\def\dsl{$L_\odot$}
\def\teff{$T_{\rm eff}$}
\def\dov{$\delta_{\rm ov}$}

\shorttitle{Asteroseismic Analysis of the CoRoT Target HD 49933}
\shortauthors{Liu et al.}

\begin{document}

\title{ASTEROSEISMIC ANALYSIS OF THE CoRoT TARGET HD 49933}

\author{Zhie Liu$^{a}$, Wuming Yang$^{a, b, 1}$
\footnotetext[1]{Corresponding author: Wuming Yang},
Shaolan Bi$^{a}$, Zhijia Tian$^{a}$, Kang Liu$^{a}$, Zhishuai Ge$^{a}$,
Jie Yu$^{a}$, Tanda Li$^{c}$, Xiaoyan Tan$^{b}$, Xin He$^{a}$, Yaqian Wu$^{a}$,
and P. Chintarungruangchai$^{a}$}
\affil{$^{a}$Department of Astronomy, Beijing Normal University,Beijing 100875, China;
$\mathrm{yangwuming@bnu.edu.cn, zhieliu@mail.bnu.edu.cn}$}
\affil{$^{b}$School of Physics and Chemistry, Henan Polytechnic
University, Jiaozuo 454000, Henan, China}
\affil{$^{c}$Key Laboratory of Solar Activity, National Astronomical Observatories,
Chinese Academy of Sciences, Beijing, 100012, China}

\begin{abstract}
The frequency ratios $r_{10}$ and $r_{01}$ of HD 49933 exhibit
an increase at high frequencies. This behavior also exists
in the ratios of other stars, which is considered to result from the low
signal-to-noise ratio and the larger linewidth at the high-frequency end
and could not be predicted by stellar models in previous work.
Our calculations show that the behavior not only can be reproduced by
stellar models, but can be predicted by asymptotic formulas of the ratios.
The frequency ratios of the Sun, too, can be reproduced well by
the asymptotic formulas. The increased behavior derives from
the fact that the gradient of mean molecular weight at the bottom
of the radiative region hinders the propagation of p-modes,
while the hindrance does not exist in the convective core.
This behavior should exist in the ratios of stars with
a large convective core. The characteristic of
the ratios at high frequencies provides a strict constraint on
stellar models and aids in determining the size of the convective core
and the extent of overshooting. Observational constraints point
to a star with $M=1.28\pm0.01$ \dsm{}, $R=1.458\pm0.005$ \dsr{},
$t=1.83\pm0.1$ Gyr, $r_{cc}=0.16\pm0.02$ \dsr{}, $\alpha=1.85\pm0.05$,
and \dov{}$=0.6\pm0.2$ for HD 49933.
\end{abstract}


\section{INTRODUCTION}


\subsection{Determination of the Convective Core}
Asteroseismology has proved to be a powerful tool for determining
the fundamental parameters of stars, probing the internal structure
of stars, and diagnosing physical processes in stellar interiors
\citep{egge05, egge06, yang07b, stel09, chri10, yang10, yang11,
yang12, silv13}. Stars with masses between about 1.1 and 1.5 \dsm{}
have a convective core during their main sequence (MS).
At the same time, however, solar-like oscillations may
be present in such stars, with a
convective core increasing in size during the initial
stage of the MS of these stars before the core begins to shrink.
The presence of the convective core brings to the fore
the question of the mixing of elements in the stellar interior
caused by the core's overshooting, which prolongs the lifetime
of the burning of core hydrogen by feeding more H-rich material
into the core. In the current theory of stellar evolution,
the overshooting of the convective core is generally described by
a free parameter \dov{}. However, the uncertainty in the mass
and extension of the convective core due to overshooting directly
affects the determination of the global parameters of the stars
by asteroseismology or other studies based on stellar evolution \citep{mazu06}.
Especially for stars with a mass of around 1.1 \dsm{}, there may or may not
exist a convective core in their interior, depending on the input
physics used in the computation of their evolutions \citep{chri10}.
Thus determining the presence of the convective core and its extension
is important for understanding the structure and evolution of stars.

There is no a direct method to determine the size of a
convective core or the extension of overshooting. Fortunately,
however, the low-$l$ p-modes can penetrate into the innermost
layers of stars, which offers an opportunity to probe the convective core.
Determining the size and extension of a convective core by
means of asteroseismology was first studied by \cite{mazu06}.
However, for most of solar-like oscillating stars, the uncertainty of
the frequencies of modes with $l=2$ and $3$ is generally larger than
that of modes with $l=0$ and $1$; and the frequencies of modes
with $l=3$ are usually not extracted. Thus it is difficult to obtain
the quantities $\theta$ and $\eta$ defined by \cite{mazu06}
or the expression suggested by \cite{cunh07}.

The small spacings
\begin{equation}
d_{10}(n)\equiv-\frac{1}{2}(-\nu_{n,0}+2\nu_{n,1}-\nu_{n+1,0})
\label{d10o}
\end{equation}
and
\begin{equation}
d_{01}(n)\equiv\frac{1}{2}(-\nu_{n,1}+2\nu_{n,0}-\nu_{n-1,1})
\label{d01o}
\end{equation}
defined by \cite{roxb93} and \cite{roxb03} can be used to diagnose
the convective core of stars. In calculation, equations (\ref{d10o})
and (\ref{d01o}) are generally rewritten as the smoother five-point
separations \citep{roxb03}. By using this tool, \cite{dehe10}
studied the solar-like pulsator HD 203608. They found that the presence
or absence of a convective core in stars can be indicated by
the small spacings and that the star HD 203608 with $M\simeq0.94$ \dsm{}
and $t\simeq6.7$ Gyr in fact has a convective core. For their part,
\cite{bran10} found that the slope of $d_{01}$ relates to the size
of the jump in the sound speed at the edge of the convective core.
\cite{deme10} studied the oscillations of $\alpha$ Centauri A ($1.105\pm0.007$ \dsm{})
and concluded that the $d_{01}$ allows one to set an upper
limit to the amount of convective-core overshooting and that the model
of $\alpha$ Centauri A with a radiative core reproduces the observed
$d_{01}/\Delta\nu$ ($r_{01}$) significantly better than the model
with a convective core. Furthermore, \cite{silv11} analyzed
the sensitivity of $r_{01}$ to the central conditions of
stars during the MS evolution. They claimed that the
presence of a convective core can be detected by $r_{01}$.
Using this asteroseismic tool, \cite{silv13} tried to detect
the convective core of Perky (KIC 6106415) and Dushera (KIC 12009504).
They found that the mass of Perky is $1.11\pm0.05$ and
that of Dushera is $1.15\pm0.04$ \dsm{}, and concluded
that a convective core and core overshooting exist in Dushera,
but could not determine whether a convective core exists in Perky.

\cite{silv13} noted, moreover, that the frequency ratios $r_{01}$
and $r_{10}$ of Perky and Dushera have a sudden increase at high
frequencies. They argued that this behavior results from the low
signal-to-noise ratio and the larger line width at the high-frequency
end and is not predicted by models. Thus they did not consider the
frequency ratios on this regime in comparison with stellar models.
If this behavior derives from the stellar interior structure
and can be predicted by stellar models and pulsation theory, stellar
models can be more strictly constrained by the ratios.

\subsection{Research on HD 49933}

The frequencies of p-modes of dozens of MS stars have been
extracted \citep{appo12}. The frequencies of p-modes
of HD 49933, which is an F5V MS star, have been determined
by many authors \citep{moss05, appo08, beno09a, beno09b, kall10}.
Combing asteroseismical and non-asteroseismical data,
several authors \citep{piau09, kall10, beno10, cree11,
bigo11} have investigated various properties of HD 49933.
\cite{piau09} concluded that the effect of rotation on HD 49933 is very weak,
but that the diffusion of elements is important. \cite{beno10}
concluded that the overshooting of the convective core is needed, but
that the microscopic diffusion may be unimportant. It seems to
be difficult to reach a consensus about the mass of HD 49933.
The mass determined by \cite{piau09} is about 1.17 \dsm{}; by
\cite{cree11} 1.1-1.2 \dsm{}; by \cite{beno10} 1.19 $\pm$ 0.08 \dsm{};
by \cite{bigo11} similarly 1.20 $\pm$ 0.08 \dsm{};
but by \cite{kall10} about 1.32 \dsm{}. The difference in mass
determinations may result from whether consideration is given to
the effects of core overshooting or microscopic diffusion.
In any case, depending on the difference of mass,
differences in other parameters obtained by these authors obviously exist.

Because HD 49933 may be more massive than HD 203608, $\alpha$ Cen A,
Perky, and Dushera, it could have a convective core.
Moreover, the increase in frequency ratios
$r_{01}$ and $r_{10}$ at high frequencies also exists in it,
making it a good target for determining the size of a convective core
by means of asteroseismology and for studying the effects
of a convective core on frequency ratios. Furthermore, there are
the frequency shifts of its low-degree p-modes \citep{sala11},
which are understood arising from the effect
of magnetic activity just as that found in the Sun.
Hence, to determine the accurate stellar parameters of
HD 49933 can be significant for understanding
the structure and evolution of stars.

In present work, we focus mainly on determining accurate stellar parameters
of HD 49933 and examining whether the increase in the frequency ratios
at high frequencies can be predicted by pulsation theory and stellar models.
In Section 2, we introduce our stellar models and perform a classical asteroseismic
analysis. In Section 3, we show the calculated results; and in Section 4,
we discuss and summarize our results.

\section{STELLAR MODELS}

\subsection{Non-asteroseismic Observational Constraints}
The luminosity of HD 49933 is 3.58 $\pm$ 0.1 \dsl{} \citep{mich08, kall10},
and its radius is $1.42 \pm 0.04$ \dsr{} \citep{bigo11}.
Its effective temperature is between about 6467 and 6780 K \citep{gill06}.
Combining other data, we took an average effective temperature:
\teff{} $=6647\pm116$ K. HD 49933 is a slightly metal-poor star
with [Fe/H] between about -0.2 and -0.5 \citep{gill06}.
For Population I stars, the ratio of surface heavy-element
abundance to hydrogen abundance is related to
the Fe/H by [Fe/H] $=\log(Z/X)_{s} - \log(Z/X)_{\odot}$,
where $(Z/X)_{\odot}$ is the ratio of the solar mixture.
The most recent ratio of the heavy-element
abundance to hydrogen abundance of the Sun, $(Z/X)_{\odot}$, is
0.0171 \citep{aspl04}. There are, however, some discrepancies between
the solar model constructed according to this new value
and seismical results \citep{yang07a}. The old value of
$(Z/X)_{\odot}$ is 0.023 \citep{grev98}, which is in good agreement
with the seismical results. The range of $(Z/X)_{s}$ for HD 49933
can be considered to be approximately between $0.0054$ and $0.0145$,
i.e. $(Z/X)_{s}\approx0.010\pm0.005$. Considering the controversy
about the value of $(Z/X)_{\odot}$, the value of $(Z/X)_{s}$
in our calculations is used as a reference, but not applied to
constraining stellar models.

In order to reproduce observed characteristics of HD 49933, we computed
a grid of evolutionary tracks using the Yale Rotation Evolution
Code \citep{pins89, guen92, yang07a}. The OPAL EOS tables \citep{roge02},
OPAL opacity tables \citep{igle96}, and the opacity tables for low
temperature provided by \cite{ferg05} were used. Energy transfer
by convection is treated according to the standard mixing length theory.
The diffusion of both helium and heavy elements is computed by
using the diffusion coefficients of \cite{thou94}. With the constraints
mentioned above, we obtained hundreds of evolutionary tracks for HD 49933.
The initial parameters of these tracks are summarized in Table \ref{para1}.

\subsection{Asteroseismic Observational Constraints}

The low-degree p-modes of HD 49933 were first identified by \cite{appo08}
and \cite{beno09a}. Using longer CoRoT \citep{bagl06} timeseries data
of HD 49933, \cite{beno09b} extracted the frequencies of the low-degree
p-modes with less uncertainty. Nonetheless, the frequencies of the modes
with $l=2$ still have a large error ($\sigma\gtrsim 2$ $\mu$Hz)
due to the fact that these modes overlap with the $l=0$ modes
and that the rotational splitting is relatively large for HD 49933
\citep{beno09b}. Here we thus only used the frequencies of
the modes with $l = 0$ and $1$ to constrain our stellar models.
For each observed frequency, we took the maximum error listed
by \cite{beno09b} at $1\sigma$. The average large frequency separation
$\overline{\Delta\nu}$ of these modes is $85.99\pm0.33$ $\mu$Hz.

\begin{figure}
\centering
\includegraphics[scale=0.38, angle=0]{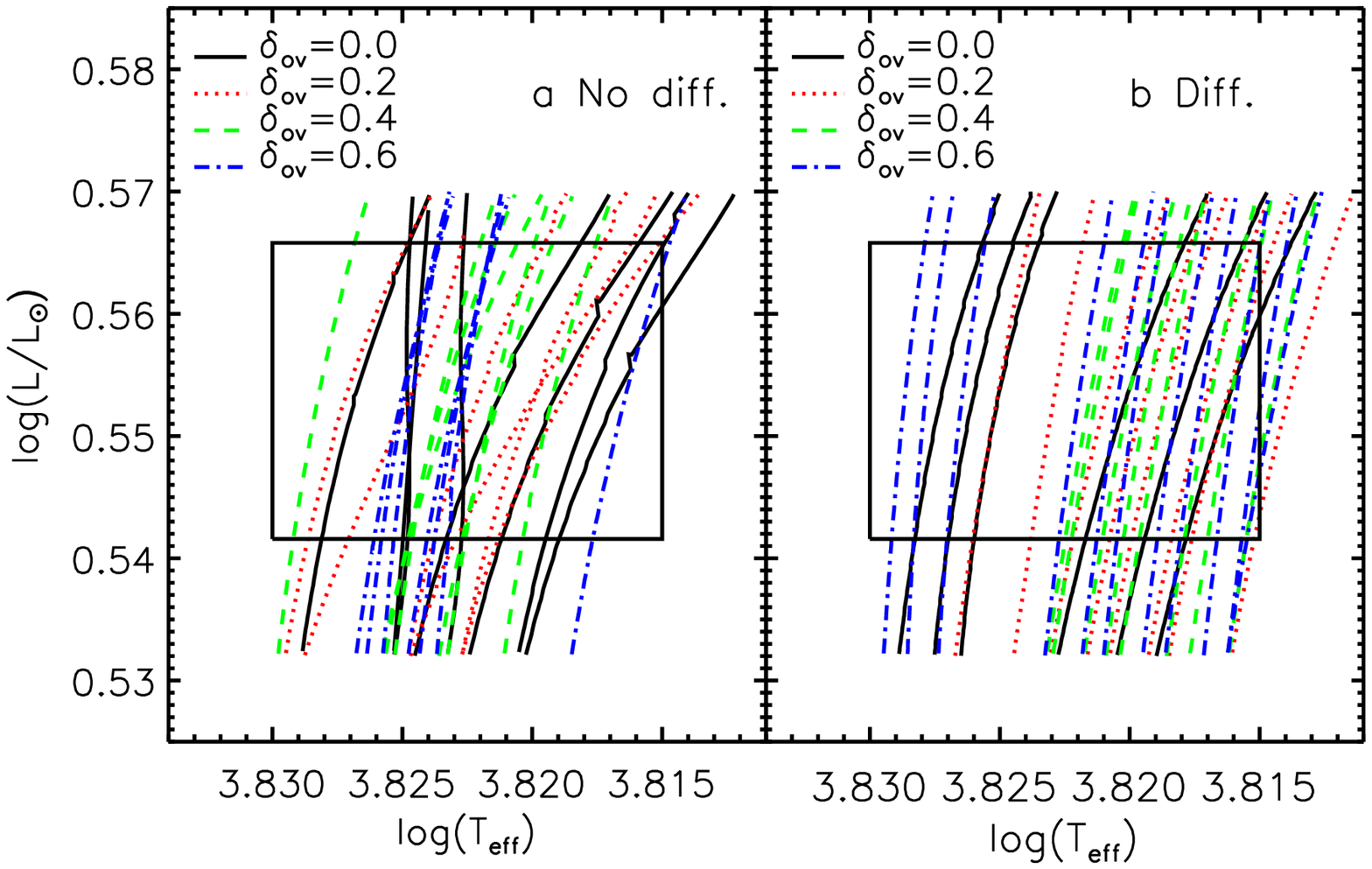}
\caption{The Hertzsprung-Russell (H-R) diagrams of models for HD 49933.
Panel $a$ shows the tracks of models without diffusion.
Panel $b$ depicts the tracks of models with diffusion.}
\label{fhrd1}
\end{figure}

In order to find the set of modeling parameters ($M, \alpha$,
\dov{}, $Z_{i},X_{i},t$) that leads to the best agreement with
the observational constraints, we computed the value of
$\chi_{c}^{2}$ of models on the grid.
The function $\chi_{c}^{2}$ is defined as follows
\begin{equation}
\chi^{2}_{c} = \frac{1}{4}\sum_{i=1}^{4}
(\frac{C_{i}^{theo}-C_{i}^{obs}}{\sigma C_{i}^{obs}})^{2},
\end{equation}
where the quantity $C_{i}^{obs}$ and $C_{i}^{theo}$ are the observed
and model values of \teff{}, $L/L_{\odot}$, $R/R_{\odot}$,
and $\overline{\Delta\nu}$, respectively.
The observational uncertainty is indicated by $\sigma C_{i}^{obs}$.
We obtained more than fifty tracks which are shown in
Figure \ref{fhrd1}. The parameters of the models on these tracks can
meet $\chi_{c}^{2}<1$. Then we computed the frequencies of
low-degree p-modes of the models on these tracks
using the Guenther adiabatic pulsation code \citep{guen94},
and we calculated the value of $\chi_{\nu}^{2}$ of each model.
The function $\chi_{\nu}^{2}$ is defined as follows
\begin{equation}
\chi_{\nu}^{2} = \frac{1}{N}\sum_{i=1}^{N}
(\frac{\nu_{i}^{theo}-\nu_{i}^{obs}}{\sigma\nu_{i}^{obs}})^{2},
\end{equation}
where $\nu_{i}^{obs}$ and $\nu_{i}^{theo}$ are the observed and
corresponding model eigenfrequencies of the $i$th mode, respectively,
and $\sigma\nu_{i}^{obs}$ is the observational uncertainty of
the $i$th mode.

\begin{figure}
\centering
\includegraphics[scale=0.35, angle=0]{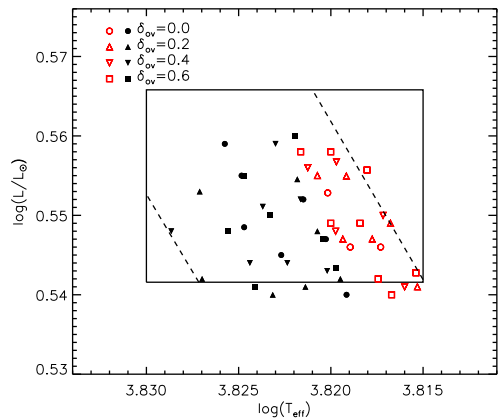}
\caption{The locations of 50 models listed in Tables \ref{tmod1} and \ref{tmod2}
in the H-R diagram. The dashed lines show the radius given by \cite{bigo11}.
Filled black symbols represent models without diffusion, while open red ones
indicate models with diffusion.}
\label{fhrd2}
\end{figure}

\begin{figure}
\centering
\includegraphics[scale=0.38]{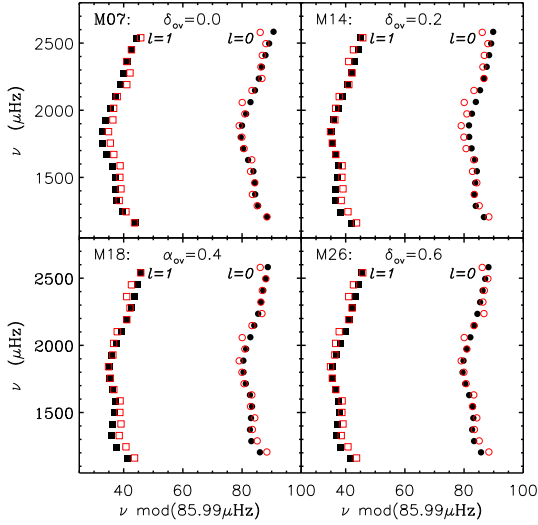}
\caption{Echelle diagrams of models without diffusion. Filled black symbols
refer to the observed frequencies, while open red ones correspond to
the theoretical frequencies.}
\label{fec1}
\end{figure}

\begin{figure}
\centering
\includegraphics[scale=0.35]{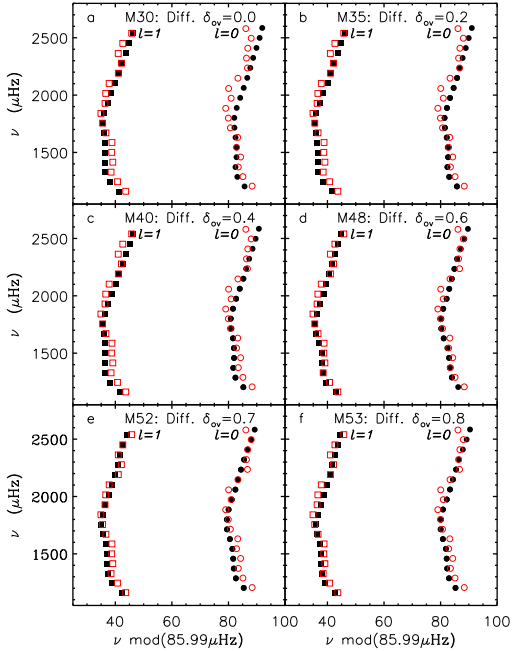}
\caption{Same as Figure \ref{fec1} but for models with diffusion.}
\label{fec2}
\end{figure}

In order to ensure that the model with the minimum
$\chi_{\nu}^{2}$ is picked out as a suitable candidate,
the time-step of the evolution for each track is set as small as
possible when the model evolves to the vicinity of the error-box
of luminosity and effective temperature in the H-R diagram. This makes
the consecutive models have an approximately equal $\chi_{\nu}^{2}$.
As a result, we obtained more than 50 models which are listed
in Tables \ref{tmod1} and \ref{tmod2}, classified according to
whether the diffusion of elements is considered or not.
Their positions in the H-R diagram are shown in Figure \ref{fhrd2}.
For the diffusion or non-diffusion models with a given \dov{},
the model with a minimum $\chi_{\nu}^{2}$ is chosen
as the candidate that best fits the model.
We thus obtained four models (M07, M14, M18, and M26)
in which element diffusion is not considered and
four other models (M30, M35, M40, and M48) in which
it is considered. The echelle diagrams of these models are
shown in Figures \ref{fec1} and \ref{fec2}.

Tables \ref{tmod1} and \ref{tmod2} show that the values
of $\chi_{\nu}^{2}$ of the models M07, M14, M18, and M26
are generally smaller than those of M30, M35, M40,
and M48. By comparing individual theoretical frequencies to
the observed ones, the method of the $\chi_{\nu}^{2}$ minimization
seems to indicate that model M26 is the one best fit for
HD 49933 because it has the minimum $\chi_{\nu}^{2}$. The echelle
diagram of model M26 is also slightly better than those of other models.
This seems to suggest that the overshooting of the convective core
is needed for HD 49933, but that the microscopic diffusion may be unimportant.
These results are consistent with those obtained by \cite{beno10}.
In addition, the mass of M26 is 1.20 \dsm{} which is also in agreement with
the value of $1.19\pm0.08$ \dsm{} determined by \cite{beno10}.
However, \cite{dehe10} and \cite{silv13} have pointed out that the
most fitting model as found solely by matching individual frequencies
of oscillations does not always properly reproduce the ratios
$r_{01}$ and $r_{10}$, i.e., the interior structure of the ``best-fit model''
may not match the interior structure of the star.

\subsection{The Effect of Near-surface Correction on $\chi_{\nu}^{2}$}
There is a well-known near-surface effect in heliosismology.
A similar effect must occur as well in asteroseismology \citep{kjel08},
which could affect results when one compares theoretical individual
frequencies to observed ones. An empirical formula was proposed
by \cite{kjel08} to correct the near-surface effect for models
of solar-like stars. When we calculated the values of
$\chi_{\nu}^{2}$ listed in Tables \ref{tmod1} and \ref{tmod2},
such a correction was not taken into account.

In order to understand the effect of the surface correction on
\textbf{all computed} models, we calculated \textbf{the chi squared
after frequency correction ($\chi^{2}_{\nu_\mathrm{corr}}$)},
using equations (6) and (2) of \cite{kjel08}.
\textbf{For a given \dov{}, the surface correction does not change
the ``best candidate" (see Table \ref{tmod1} and \ref{tmod2}).
However in overall and for models without diffusion,
while a model with a 0.2 overshoot was favored (M14),
a model with no overshoot is now favored (M07).
Nonetheless, according to the theory of \cite{kjel08},
one should also account of the correction factor $r_{f}$ to define
the best model. The closer a reference model is to the best one,
the closer to $1$ is $r_{f}$. In other words, the uncorrected and
corrected chi squared must be the same and small.
The value of the correction factor $r_{f}$ is $1.00031$ for M14,
which among all our models, is the closest to one. For models
with diffusion, we favor M48 because of its $\chi^{2}_{\nu_\mathrm{corr}}$
and because $r_{f}=1.00038$. We therefore conclude that M14 and M48
are the best models that fit the observations.}

\section{ASTEROSEISMICAL DIAGNOSIS}
\subsection{The Frequency Ratios of Models}

\begin{figure}
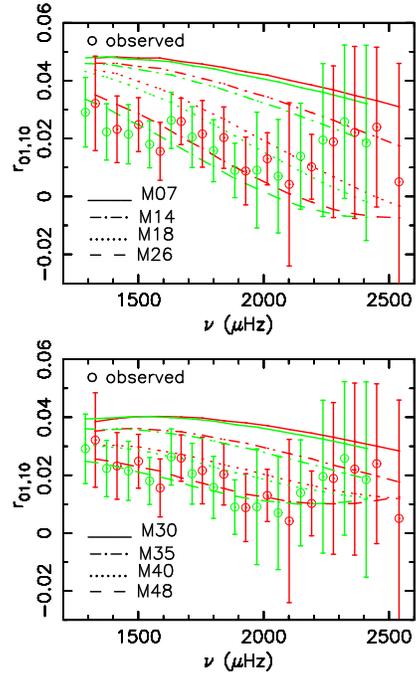

\centering
\includegraphics[scale=0.5, angle=-90]{fig5-1.ps}
\includegraphics[scale=0.5, angle=-90]{fig5-2.ps}
\caption{The ratios $r_{01}$ and $r_{10}$ of observational frequencies
and adiabatic frequencies of models. The red shows $r_{01}$.
The green represents $r_{10}$. For each frequency, we accepted
the maximum error bar listed by \cite{beno09b} at $1\sigma$. }
\label{fd011}
\end{figure}

We calculated the ratios $r_{01}$ and $r_{10}$ of the observed and theoretical
frequencies. Figure \ref{fd011} shows that the observed $r_{01}$ and $r_{10}$
decrease with a frequency in the approximate range between about 1300 and
2050 $\mu$Hz (hereafter labeled ``decrease range"), but they increase with
a frequency in the range between about 2050 and 2400 $\mu$Hz (labeled
``increase range"). The model M14 cannot reproduce the observed
ratios $r_{01}$ and $r_{10}$. However, models M26, M40, and M48 reproduce
well the ratios in the decrease range. The same ratios
of models M26 and M40 decrease in the increase range, which is inconsistent
with the observed results; but those of model M48 increase slightly in
the range between about 2100 and 2500 $\mu$Hz. Although the ratios of M48
cannot reproduce the increase at high frequencies, the trend of increase
is consistent with the behavior of the observed ratios. Furthermore,
Figure \ref{fd011} shows that the observed $r_{01}$ is slightly larger
than $r_{10}$ in the decrease range, but smaller in the increase range.
The ratios of M48 exhibit the same characteristics.

\begin{figure}
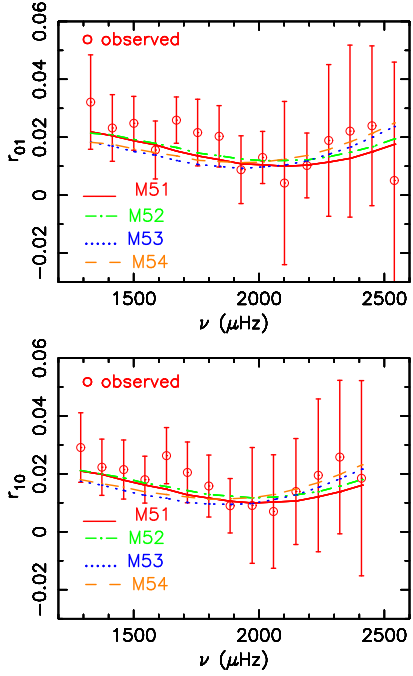

\centering
\includegraphics[scale=0.5, angle=-90]{fig6-1.ps}
\includegraphics[scale=0.5, angle=-90]{fig6-2.ps}
\caption{Same as the Figure \ref{fd011}, but for different models. }
\label{fd012}
\end{figure}

Models M26 and M48 have a large \dov{}. The increase behavior of
the frequency ratios at high frequencies might be related to the
convective core, which can be significantly affected by overshooting
parameter \dov{}. Thus we constructed four models with a
larger \dov{}, i.e. M51, M52, M53, and M54. These
reproduce not only the ratios in the decrease range,
 but also the increase behavior of the ratios at high frequencies
(see Figure \ref{fd012}). In addition, the ratio $r_{01}$ of these
models is larger than their $r_{10}$ at low frequencies, but
smaller at high frequencies. The increase in the ratios
$r_{01}$ and $r_{10}$ for these models is more obvious at high
frequencies. The behavior of the increase could derive from
the interior structure of stars.

\subsection{Asymptotic Formula and Modification}
\subsubsection{Asymptotic Formula of Frequency Ratios}
In order to better understand whether the increase behavior results
from the structure of stars or not, we analyze the characteristics
of the ratios by making use of the asymptotic formula of frequencies.
That for the frequency $\nu_{n,l}$ of a stellar p-mode of
order \textit{n} and degree \textit{l} is given by \cite{tass80}
and \cite{goug90} as
\begin{equation}
\nu_{n,l} \simeq (n+\frac{l}{2}+\varepsilon)\nu_0-[Al(l+1)-B]\nu_0^2\nu_{n,l}^{-1}
\label{eqnu}
\end{equation}
for $n/(l+\frac{1}{2})\rightarrow\infty$, where
\begin{equation}
\nu_0=(2\int_0^R\frac{dr}{c})^{-1}
\end{equation}
 and
\begin{equation}
A=\frac{1}{4\pi^2\nu_0}[\frac{c(R)}{R}-\int_{r_{t}}^{R}\frac{1}{r}\frac{dc}{dr}dr].
\end{equation}
In this equation \textit{c} is the adiabatic sound speed at
radius \textit{r} and \textit{R} is the fiducial radius of the star;
$\varepsilon$ and \textit{B} are the quantities that
are independent of the mode of oscillation, but depend predominantly
on the structure of the outer parts of the star; $r_{t}$ is the inner
turning point of the mode with the frequency $\nu_{n,l}$ and
can be determined by
\begin{equation}
\frac{c(r_{t})}{r_{t}}=\frac{2\pi\nu_{n,l}}{\sqrt{l(l+1)}};
\label{rt}
\end{equation}
$\nu_{0}$ is related to the travel time of sound across
the stellar diameter; $A$ is a measure of the sound-speed gradient
and is most sensitive to conditions in the stellar core and
to changes in the composition profile \citep{goug90, chri93}.

Using this asymptotic formula, one can obtain
\begin{equation}
d_{10}(n)=\frac{1}{2}[4A\nu_{0}^{2}\nu_{n,1}^{-1}-B\nu_{0}^{2}
(2\nu_{n,1}^{-1}-\nu_{n,0}^{-1}-\nu_{n+1,0}^{-1})]
\end{equation}
and
\begin{equation}
d_{01}(n)=\frac{1}{2}[2A\nu_{0}^{2}(\nu_{n,1}^{-1}+\nu_{n-1,1}^{-1})
+B\nu_{0}^{2}(2\nu_{n,0}^{-1}-\nu_{n,1}^{-1}-\nu_{n-1,1}^{-1})].
\end{equation}
The quantities $A$ and $B$ have the same order [see equation (72) of \cite{tass80}].
Because $2\nu_{n,1}^{-1}\gg |2\nu_{n,1}^{-1}-\nu_{n,0}^{-1}-\nu_{n+1,0}^{-1}|$
and $\nu_{n,1}^{-1}+\nu_{n-1,1}^{-1}\gg |2\nu_{n,0}^{-1}-\nu_{n,1}^{-1}-\nu_{n-1,1}^{-1}|$,
then
\begin{equation}
d_{10}(n)\simeq A\nu_{0}^{2}2\nu_{n,1}^{-1}
\label{d10}
\end{equation}
and
\begin{equation}
d_{01}(n)\simeq A\nu_{0}^{2}(\nu_{n,1}^{-1}+\nu_{n-1,1}^{-1}).
\label{d01}
\end{equation}
Using $\Delta\nu\simeq\nu_{0}$, we can get
\begin{equation}
r_{10}(n)\equiv\frac{d_{10}(n)}{\Delta\nu}\simeq A\nu_{0}2\nu_{n,1}^{-1}
\label{r10}
\end{equation}
and
\begin{equation}
r_{01}(n)\equiv\frac{d_{01}(n)}{\Delta\nu}\simeq A\nu_{0}(\nu_{n,1}^{-1}+\nu_{n-1,1}^{-1}),
\label{r01}
\end{equation}
where
\begin{equation}
A\nu_{0}=\frac{1}{4\pi^2}[\frac{c(R)}{R}+\int_{r_{t}}^{R}(-\frac{1}{r}\frac{dc}{dr})dr].
\label{av0}
\end{equation}
The quantity $1/\nu_{n,1}$ decreases monotonically with the increase in frequencies.
Thus, the sudden change in the ratios could result from the change in $A\nu_{0}$.

\subsubsection{Characteristics Predicted by the Asymptotic Formula}
\begin{figure}
\centering
\includegraphics[scale=0.5, angle=-90]{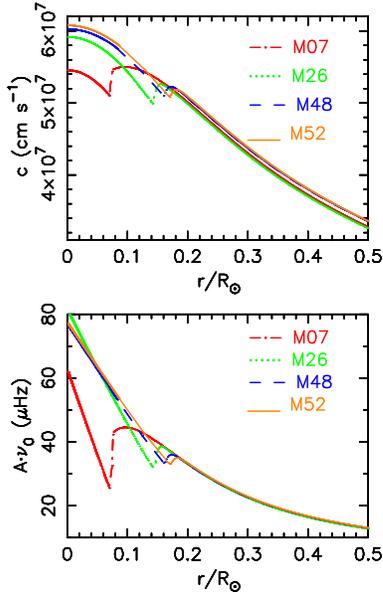}
\caption{The radial distributions of adiabatic sound speed
and quantity $A\nu_{0}$ of different models.}
\label{fcr}
\end{figure}

The radial distributions of the adiabatic sound speed and the quantity $A\nu_{0}$ of
different models are shown in Figure \ref{fcr}. The quantity $A\nu_{0}$ increases with
the decrease in radius and depends mainly on the conditions in the stellar core.
According to equation (\ref{rt}), the higher the frequencies, the smaller the $r_{t}$
for modes with a given $l$. Thus the quantity $A\nu_{0}$ increases with
increasing frequencies. When the decrease of $2/\nu_{n,1}$ cannot counteract
the increase of $A\nu_{0}$, the ratio $r_{10}$ would increase with increasing
frequencies.

Moreover, if changes in the ratios $r_{10}$ and $r_{01}$ are dominated mainly
by the changes of $2/\nu_{n,1}$ and ($1/\nu_{n,1}+1/\nu_{n-1,1}$)
respectively, equations (\ref{r10}) and (\ref{r01}) show that the ratios
decrease with increasing frequencies and that $r_{01}$ can be
slightly larger than $r_{10}$ because $(1/\nu_{n,1}+1/\nu_{n-1,1})>2/\nu_{n,1}$.
This case happens at low frequencies because the lower the frequency,
the larger the difference between $1/\nu_{n,1}$ and $1/\nu_{n-1,1}$.
However, if the changes in the ratios are mainly dominated by the changes of
$A\nu_{0}$ and the difference between $1/\nu_{n,1}$ and $1/\nu_{n-1,1}$ can be
neglected, the ratios increase with the increase in $A\nu_{0}$, and $r_{01}$
would be slightly smaller than $r_{10}$ because the $A\nu_{0}$ corresponding
to $\nu_{n-1,1}$ is smaller than that corresponding to $\nu_{n,1}$. This case
occurs at high frequencies because the higher the frequency, the smaller
the difference between $1/\nu_{n,1}$ and $1/\nu_{n-1,1}$. These characteristics
are similar to those of the ratios of the observed frequencies and the adiabatic
oscillation frequencies of models M48, M51, M52, M53, and M54.
Thus an increase could exist in the ratios of p-mode frequencies
of a star, and it could be seen within the observed frequency
range of the star, provided that the slope change is close to
the frequencies of the maximum seismical amplitude ($\nu_{max}$).

Figure \ref{fcr} shows that the quantity $A\nu_{0}$ varies significantly
with the stellar radius in the convective core, which may lead to the
possibility that the ratios $r_{10}$ and $r_{01}$ increase with increasing
frequencies. By making use of asymptotic formulas (\ref{rt}) and (\ref{r10}),
our calculations show that the ratios of M07 exhibit an insignificant
increase at high frequencies [see the dotted (blue) line in the panel M07
of Figure \ref{fr104}], which shows that the increase in ratios can be
predicted by the asymptotic formulas despite the fact that the increase
cannot match the observed one. However, the ratio of the adiabatic oscillation
frequencies of M07 (the green dash-dotted line) does not exhibit
a similar behavior.

\subsubsection{Modification to the Formula}
\begin{figure}
\centering
\includegraphics[scale=0.6, angle=-90]{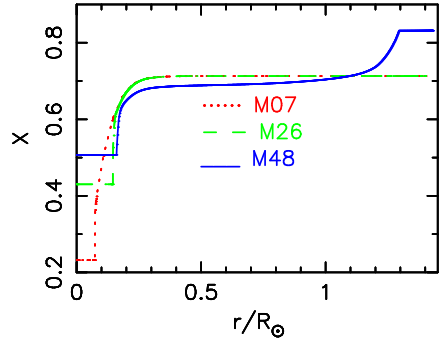}
\caption{The radial distributions of hydrogen abundance of different models.}
\label{fhy}
\end{figure}

The value of $r_{10}$ and $r_{01}$ computed using the asymptotic
formulas (\ref{r10}) and (\ref{r01}) strongly depend on the
equation (\ref{rt}), which is obtained from the dispersion relation
of p-modes. The dispersion relation is deduced from adiabatic
oscillation equations with the conditions of Cowling approximation
and local (constant coefficient) approximation \citep{unno89}.
For stars with a convective core, there is a large gradient of chemical
composition $\nabla_{\mu}$ at the top of the core (see Figure
\ref{fhy}), which leads to a sudden increase in the squared
Brunt-V$\ddot{a}$is$\ddot{a}$l$\ddot{a}$ frequency $N^{2}$.
This results in the fact that the coefficient
\{$h(r)=\exp[\int^{r}_{0}(N^{2}/g-g/c^{2})dr]$\} of the adiabatic oscillation
equations [(15.5) and (15.6) in \cite{unno89}] strongly depends on
radius $r$ in this region; i.e. the coefficients should not be assumed
to be constant. Thus relation (\ref{rt}) should be modified. Moreover,
the large $\nabla_{\mu}$ might hinder the fact that p-modes
transmit in the stellar interior. In other words,
the equation (\ref{rt}) might underestimate the value of
the turning point $r_{t}$. Therefore, we modify relation (\ref{rt})
as
\begin{equation}
\nu_{n,l}=f_{0}\frac{c(r_{t})}{r_{t}}\frac{\sqrt{l(l+1)}}{2\pi},
\label{rtm}
\end{equation}
where the parameter $f_{0}$ is larger than one.
For a given mode, the position of the turning point determined by
equation (\ref{rt}) is deeper than that predicted by equation (\ref{rtm}).
Thus the frequency ratios computed by equations (\ref{r10}) and (\ref{rt})
are larger than those calculated by equations (\ref{r10}) and (\ref{rtm}).

\subsection{Calculated Results of Asymptotic Formula}
\subsubsection{Results of HD 49933}
\begin{figure}
\centering
\includegraphics[scale=0.38, angle=-90]{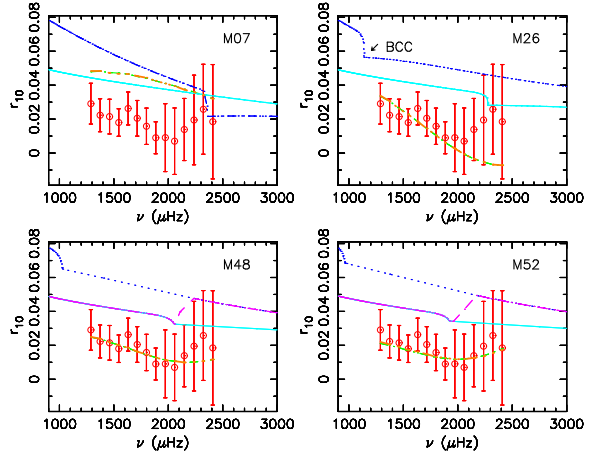}
\caption {The ratio $r_{10}$ as a function of oscillation frequencies.
The circles with an error bar show the observed data. The dash-dotted
green line indicates the ratio of adiabatic oscillation frequencies.
The dotted blue line shows the ratio computed from equations (\ref{r10})
and (\ref{rt}). The solid cyan line depicts the ratio computed using
equations (\ref{r10}) and (\ref{rtm}). The long-dashed magenta line
represents the one computed using equations (\ref{r10}) and (\ref{rtm})
in the radiative region, but computed using equations (\ref{r10}) and (\ref{rt})
in the convective core, while the overshoot region has been treated as a
transition region. The dash-triple-dotted orange line depicts the ratio
for the corrected adiabatic oscillation frequencies. It overlaps
with the uncorrected one.
The BCC shows the boundary of the convective core.}
\label{fr104}
\end{figure}

The observed $r_{10}$ has an increase at about 2050 $\mu$Hz.
If the increase is related to the convective core,
we can use the asymptotic formulas (\ref{r10}) and (\ref{rtm})
and the structure of model M48 to calibrate equation (\ref{rtm})
and determine that the value of $f_{0}$ is 2.0.
The solid (cyan) lines in Figure \ref{fr104} show the computed $r_{10}$
using formulas (\ref{r10}) and (\ref{rtm}). For M07,
equation (\ref{rtm}) predicts that modes with a frequency
between 1000 and 2500 $\mu$Hz cannot reach the convective core.
Thus the ratio $r_{10}$ of adiabatic oscillation frequencies
in this range does not exhibit increased behavior. For M26
and M52, Figure \ref{fr104} shows that the ratio $r_{10}$ of adiabatic
oscillation frequencies starts to increase at about 2400 and 2000 $\mu$Hz
respectively. Formulas (\ref{r10}) and (\ref{rtm}) predict that
modes with a frequency larger than about 2300 $\mu$Hz
for M26 and 1950 $\mu$Hz for M52 can reach the
convective core. Unfortunately, the ratio $r_{10}$ determined
by formulas (\ref{r10}) and (\ref{rtm}) does not reproduce
directly the observed increase behavior.

\subsubsection{Further Assumption and Results}
In a convective region, there is no gradient of chemical composition;
so the turning point should be determined by equation (\ref{rt}).
Due to the fact that the quantity $A\nu_{0}$ is mainly dependent
on the conditions in stellar interiors, we assume that the
turning point of the modes propagating out of the convective core
is determined by equation (\ref{rtm}) despite the fact that
there is no chemical gradient in the outer layers of the radiative region,
but that the turning point of the modes penetrated the convective core
is determined by equation (\ref{rt}). These assumptions can result
in two inferences. One is that the frequency ratios
can increase suddenly near the frequencies
whose corresponding turning points are located at the boundary
of the convective core. The other is that the magnitude of the increase
is equal to the difference between the ratios determined by
equations (\ref{r10}) and (\ref{rt}) and those determined by
equations (\ref{r10}) and (\ref{rtm}) at the boundary of the convective core.

For model M48, according to formula (\ref{rtm}), the turning point
of the modes of about 2100 $\mu$Hz is located at the boundary
of the convective core. For the modes with frequencies larger
than 2100 $\mu$Hz, their turning points should be determined by equation
(\ref{rt}) in the convective core. The value of $r_{10}$ determined
by formulas (\ref{r10}) and (\ref{rt}) is about 0.0494 at around
2100 $\mu$Hz, while that determined by formulas (\ref{r10}) and (\ref{rtm})
is about 0.0323 for the frequency just less than 2100 $\mu$Hz.
The difference between these two values is about 0.017, which is
compatible with the value of 0.019 of the increase of the observed
$r_{10}$ at around 2100 $\mu$Hz. A sudden change in the turning point
of the consecutive $\nu_{n,1}$ seems unreasonable. Thus we assume
that the overshooting region is a transition region between the radiative region
and the convective core as determined by Schwarzschild criteria;
that is to say, the value of $f_{0}$ changes from 2.0 to 1.0 in
the overshooting region. The ratios $r_{10}$ computed according
to this assumption are shown by the long-dashed (magenta) lines
in Figure \ref{fr104}. The magnitude of the increase of the $r_{10}$
between about 2050 and 2250 $\mu$Hz is about 0.015 for M48,
which is slightly lower than the observed value of 0.019.
For M52, when the overshooting region is treated as
a transition region, the magnitude of the increase
in the $r_{10}$ between 1950 and 2150 $\mu$Hz is also about 0.015.

\subsubsection{Test by the Sun}
\begin{figure}
\centering
\includegraphics[scale=0.6, angle=-90]{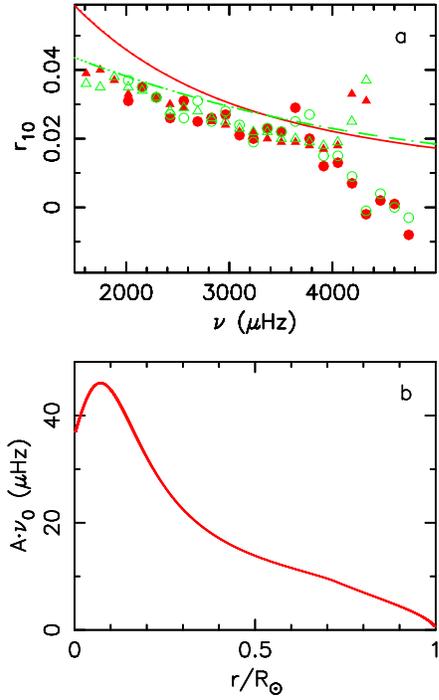}
\caption {Panel $a$: The solid red and open green circles show
the $r_{01}$ and $r_{10}$ of the Sun \citep{duva88}, while
the $r_{01}$ and $r_{10}$ of the Sun \citep{garc11} are indicated by
the solid red and open green triangles, respectively. The solid red
line represents the ratio $r_{10}$ computed using solar model M98
\citep{yang07a} and equations (\ref{r10}) and (\ref{rt});
the dashed green line depicts the ratio $r_{10}$ computed
using solar model M98 and equations (\ref{r10}) and (\ref{rtm}).
Panel $b$ indicates the distribution of $A\nu_{0}$ of model M98.
The turning point of 4000 $\mu$Hz is located at about 0.082 \dsr{}.}
\label{fsun}
\end{figure}

There exists a gradient of the mean molecular weight in the interior
of the Sun despite the fact that it is less than the gradient in
stars with a convective core. Thus the calibrated equation (\ref{rtm})
can be tested by the solar model and solar oscillation frequencies.
We computed the ratios $r_{10}$ and $r_{01}$ of the Sun
from the frequencies given by \cite{duva88} and \cite{garc11}
(see the panel $a$ of Figure \ref{fsun}) respectively,
and the values of the asymptotic $r_{10}$ for the solar
model M98 \citep{yang07a}. The solid (red) line in the panel $a$ of
Figure \ref{fsun} shows the $r_{10}$ computed using equations (\ref{r10})
and (\ref{rt}). The dashed (green) line represents the $r_{10}$
computed using equations (\ref{r10}) and (\ref{rtm}), which
is in good agreement with the observed $r_{10}$ between about 2000 and
3700 $\mu$Hz for the frequencies calculated by \cite{duva88}
or between about 1500 and 4000 $\mu$Hz for those calculated
by \cite{garc11}.

However, the asymptotic formulas do not reproduce
the observed $r_{10}$ when frequencies are larger
than 4000 $\mu$Hz. In this range, the value of the
asymptotic $r_{10}$ is larger than that obtained from the
frequencies of \cite{duva88}, but lower than that obtained
from those of \cite{garc11}. According to equation (\ref{rtm}),
the turning point $r_{t}$ of the modes of 4000 and 4800 $\mu$Hz
is located at about 0.082 and 0.069 \dsr{},
respectively. Panel $b$ of Figure \ref{fsun} shows that
the quantity $A\nu_{0}$ reaches the maximum at about 0.07 \dsr{}.
Then the quantity decreases with decreasing radius; i.e.,
$A\nu_{0}$ decreases with the increase in frequencies when
the frequencies are larger than a certain value, which
leads to the fact that the ratios of the modes propagating
in this region could decrease more rapidly with the increase
in frequencies. The discrepancy between the observed
$r_{10}$ and the asymptotic $r_{10}$ at high frequencies
could reflect that between the central structure of
the solar model and that of the Sun.

\subsection{The Influence of Near-surface Effects on the Ratios}
In principle, the ratios $r_{10}$ and $r_{01}$ should not
be sensitive to surface effects. In order to determine whether
they are actually affected by such effects, we computed the
ratios of the corrected frequencies of our all models.
The ratio $r_{10}$ of the corrected frequencies of models M07,
M26, M48, and M52 is shown in Figure \ref{fr104} by the
dash-triple-dotted (orange) line, which
completely overlaps with the uncorrected one. Our calculations show
that the ratios $r_{10}$ and $r_{01}$ are not affected by such effects.

\section{DISCUSSIONS AND SUMMARY}
\subsection{Discussions}
\cite{silv11} argued that the slope of the ratios $r_{10}$ and $r_{01}$
depends simultaneously on the central hydrogen content \textbf{($X_{c}$)},
the extent of
the convective core, and the amplitude of the sound-speed discontinuity
at the core boundary. \cite{bran10} also concluded that the relation
between the slope of $d_{01}$ and the size of the jump in the sound
speed at the edge of the convective core is linear. In our models,
however, the slope seems not to depend directly on the central hydrogen
content, the extent of the convective core, or the size of
the jump in the sound speed. Model M26 has almost the steepest
ratios in the observable range of frequencies (see Figure \ref{fr104}).
However, its central hydrogen content and the size of its convective
core are larger than those of model M07, though less than
those of M48. At the same time the amplitude of the
sound-speed discontinuity at the core boundary (see Figure \ref{fcr})
is less than that of M07, but larger than that of M48.

According to equation (\ref{r10}), due to the fact that $2/\nu_{n,1}$
decreases with increasing frequencies, the slope of ratio $r_{10}$
is mainly affected by $A\nu_{0}$. If a mode did not
arrive at the convective core, its $A\nu_{0}$
could not be affected by that core.
According to equation (\ref{rtm}), the value of $A\nu_{0}$
is about 28.7 $\mu$Hz at 1300 $\mu$Hz for models M07 and M26;
but that value is about 40.7 $\mu$Hz at 2400 $\mu$Hz, whose
corresponding turning point is located in the radiative
region for M07. Thus the $r_{10}$ of model M07 only slightly decreases with
the increase in frequencies and is not affected by the jump in the
sound speed at the core boundary. For model M26, the value is 32.6 $\mu$Hz
at 2300 $\mu$Hz, whose corresponding turning point is located
just at the core boundary; in this case the jump in the sound speed
at the core boundary affects the slope of $r_{10}$ of adiabatic frequencies.

The overshooting region is treated as a transition area between the
radiative region and the convective core, which means that the
mixing should not be a complete process in the overshooting region.
This is consistent with the result that overshooting mixing should be
regarded as a weak mixing process in stars \citep{zhan12, zhan13}.

The distribution of $r_{10}$ obtained from the asymptotic formula
has a shift, compared to the distributions of the observed $r_{10}$
(see Figure \ref{fr104}). This may suggest that there
are some discrepancies between our models and HD 49933.

Moreover, the uncertainty of the observed ratios is relatively
large at high frequencies, which is mainly due to the uncertainty of the
$l=0$ modes. Considering the large uncertainty, the increase of the ratios
is not very significant. This may affect our interpretation on HD 49933.

In our calculations, only the models with a large convective core
can reproduce the increase in the ratios $r_{10}$ and $r_{01}$.
For these models, the value of \dov{} is around 0.6, which
is larger than the generally accepted value. This may be why the
behavior cannot be predicted by the models of \cite{silv13}.
Moreover, the surface rotation period of HD 49933 is about 3.4 days
\citep{beno09b}, which is obviously lower than the approximately
27 days of the Sun. Rotation can lead to an increase in the convective core,
which depends on the rotational mixing and rotation rate \citep{maed87, yang13a}.
Thus, the large core might be related to the rotation. In addition,
the low-mass stars with mass more massive than about 1.25 \dsm{}
have a shallow convective zone. Their angular momentum loss rate
could be either lower than that of the Sun or
neglected \citep{ekst12, yang13a, yang13b}. The rotation rate
of such stars can obviously be higher than that of the Sun.
The high rotation rate of HD 49933, therefore, suggests that its mass
might be greater than 1.25 \dsm{}.

\subsection{Summary}
In this work, we constructed a series of models for HD49933.
The increase behavior in the ratios $r_{10}$ and $r_{01}$
at high frequencies is not only reproduced by our models,
but also predicted by the asymptotic formulas of the ratios.
The increase in the ratios could exist for stars with
a large convective core and derive from the fact
that the gradient of chemical compositions at the bottom
of the radiative region hinders the propagation of p-modes,
while such a hindrance does not exist in the convective
core. The asymptotic formulas of the ratios calibrated to HD 49933
reproduce well the solar ratios between about 1500 and 4000 $\mu$Hz.
At higher frequencies, the behavior of the ratios obtained from the
frequencies of \cite{garc11} is contrary to that obtained from the
data of \cite{duva88}; at the same time, the observed ratios cannot be
reproduced by theoretical ratios.

When the characteristic of the increase in the ratios is neglected,
the observational constraints point to a star with $M=1.20^{+0.09}_{-0.02}$ \dsm{},
$R=1.43^{+0.033}_{-0.013}$ \dsr{}, $t=2.97^{+0.06}_{-1.24}$ Gyr,
$r_{cc}=0.145^{+0.035}_{-0.023}$ \dsr{}, $\alpha=1.85\pm0.05$,
and \dov{}$=0.6\pm0.2$ for HD 49933, where the uncertainties
\textbf{correspond to the maximum/minimum values reached by
the model parameters, for $1\sigma$ departures from the most
probable value of the observational constraints.} However,
when that characteristic is used to constrain stellar models,
the observational constraints favor a star with $M=1.28\pm0.01$ \dsm{},
$R=1.458\pm0.005$ \dsr{}, $t=1.83\pm0.1$ Gyr,
$r_{cc}=0.16\pm0.02$ \dsr{}, $\alpha=1.85\pm0.05$,
and \dov{}$=0.6\pm0.2$. The characteristic of
the increase in the ratios provides a strict constraint
on stellar models and could be applied to determining the extent
of overshooting and the size of the convective core.
Moreover, in order to reproduce the characteristic of
the increase in the ratios of HD 49933, the diffusion
of chemical elements and the core overshooting are required.
However, if the frequency ratios at high frequencies are
not considered when compared to stellar models, the effect
of element diffusion can be neglected when computing the
models.

\begin{acknowledgements}
We thank the anonymous referee for helpful comments,
R. A. Garc\'{\i}a for providing the frequencies of the Sun, and
Daniel Kister for improving the English. We acknowledge
the support from the NSFC 11273012, 11273007, and 10933002,
and the Fundamental Research Funds for the Central Universities.

\end{acknowledgements}


\begin{table}
\begin{center}
\caption[]{ The initial parameters of models.}\label{para1}

\begin{tabular}{cccccc}
  \hline\noalign{\smallskip}
  \hline\noalign{\smallskip}
  &Variable     &  Minimum  & Maximum & $\delta$  \\
   \hline\noalign{\smallskip}
            &M/M$_\odot$  &  1.10     & 1.34    &  $\leq$0.02    \\
            &$\alpha$     &  1.65     & 1.95 &  $\leq$0.1     \\
            &$\delta_{\rm ov}$ &  0.0     & 0.6    &  $\leq$0.2     \\
            \hline
            Diffusion &$Z_{i}$   &  0.006    & 0.030   &  0.002   \\
            &$ X_{i} $        &  0.675    & 0.699   &  0.002   \\
            \hline
            No Diff. &$Z_{i}$          &  0.004    & 0.012   &  0.002   \\
            &$ X_{i} $        &  0.713    & 0.721   &  0.002   \\
            \hline\noalign{\smallskip}

  \noalign{\smallskip}
\end{tabular}
\end{center}
\end{table}

\clearpage

\begin{table*}
\begin{center}
\renewcommand\arraystretch{1.0}
\caption[]{Parameters of models without diffusion. The models
with the same \dov{} are treated as a set. In each set, the model with
a minimum $\chi_{\nu}^{2}$ is highlighted by italics. \textbf{The
$r_{cc}$ shows the radius of the convctive core of a model.}}
\label{tmod1}
\begin{tabular}{p{0.40cm}cccccccccccccc}
  \hline\hline\noalign{\smallskip}
   Model & $M$ & $(Z/X)_{s}$ &$X_{c}$ & \teff{}& $ L$ & $r_{cc}$& $R$& age&$\alpha$
   & \dov{} &$\Delta\nu$& $\chi_{\nu}^{2}$ & $\chi^{2}_{\nu_\mathrm{corr}}$  & $\chi_{c}^{2} $      \\
   &  (\dsm{})         & &  &      (K)  &  (\dsl{})  & (\dsr{}) & (\dsr{}) & (Gyr) & & &($\mu$Hz)& & & \\
  \hline\noalign{\smallskip}
 M1 & 1.14 & 0.0112 & 0.020 & 6679 & 3.592 & 0.042 & 1.417 & 2.978 & 1.75 & 0.0 & 86.48 & 9.47 &14.65  & 0.86\\
 M2 & 1.18 &  0.0140 & 0.079 & 6648 & 3.508 & 0.031 & 1.414 & 2.681 & 1.75 & 0.0 & 86.43 &6.86 &10.63  &0.57\\
 M3 & 1.18 &  0.0140 & 0.044 & 6681 & 3.589 & 0.078 & 1.417 & 2.764 & 1.85 & 0.0 & 86.33 &5.57 &7.49  &0.30\\
 M4 & 1.20 &  0.0140 & 0.206 & 6695 & 3.631 & 0.062 & 1.418 & 2.397 & 1.75 & 0.0 & 86.45 &10.00& 14.47 &0.59\\
 M5 & 1.22 & 0.0168 & 0.267 & 6594 & 3.467 & 0.070 & 1.429 & 2.269 & 1.65 & 0.0 & 86.45 & 9.77 &14.37  &0.86\\
 M6 & 1.22 & 0.0168 & 0.238 & 6611 & 3.524 & 0.071 & 1.433 & 2.382 & 1.75 & 0.0 & 86.27 & 8.47 & 6.92 &0.31\\
 \it{M7} & \it{1.22} & \it{0.0168} & \it{0.226} & \it{6626} & \it{3.573} & \it{0.073} & \it{1.436} & \it{2.460}
& \it{1.85} & \it{0.0} & \it{86.19} &  \it{4.56} & \it{2.75} &\it{0.14}\\
\hline
 M8 & 1.10 & 0.0083 & 0.166 & 6714 & 3.483 & 0.074 & 1.381 & 3.406 & 1.85 & 0.2 & 86.35 &   8.15 & 9.86 &0.85\\
 M9 & 1.14 & 0.0112 & 0.223 & 6655 & 3.467 & 0.083 & 1.403 & 3.125 & 1.85 & 0.2 & 86.28 &   7.82 & 7.41 &0.55\\
 M10 & 1.16 & 0.0112 & 0.286 & 6716 & 3.573 & 0.088 & 1.399 & 2.692 & 1.75 & 0.2 & 86.50 &  12.86 &16.80  &0.75\\
 M11 & 1.18 & 0.0140 & 0.294 & 6628 & 3.475 & 0.091 & 1.416 & 2.718 & 1.75 & 0.2 & 86.40 &   5.43 &9.39  &0.67\\
 M12 & 1.22 & 0.0168 & 0.364 & 6599 & 3.483 & 0.098 & 1.429 & 2.290 & 1.65 & 0.2 & 86.41 &  13.30 & 15.88 &0.69\\
 M13 & 1.22 & 0.0168 & 0.341 & 6618 & 3.532 & 0.097 & 1.433 & 2.398 & 1.75 & 0.2 & 86.39 &   5.40 & 10.28 &0.47\\
 \it{M14} & \it{1.22} & \it{0.0168} & \it{0.319} & \it{6635} & \it{3.589} & \it{0.097} & \it{1.435} & \it{2.497}
& \it{1.85} & \it{0.2} & \it{86.27} &  \it{2.50} & \it{3.00} &\it{0.23}\\
\hline
 M15 & 1.14 & 0.0112 & 0.341 & 6674 & 3.499 & 0.115 & 1.401 & 3.202 & 1.85 & 0.4 & 86.29 &   5.86 & 8.25 &0.44\\
 M16 & 1.16 & 0.0112 & 0.385 & 6726 & 3.606 & 0.120 & 1.400 & 2.776 & 1.75 & 0.4 & 86.57 &  10.84 & 21.75 &0.97\\
 M17 & 1.18 & 0.0140 & 0.390 & 6643 & 3.499 & 0.123 & 1.415 & 2.785 & 1.75 & 0.4 & 86.39 &   6.68 & 11.94 &0.53\\
 \it{M18} & \it{1.18} & \it{0.0140} & \it{0.375} & \it{6663} & \it{3.556} & \it{0.122} & \it{1.417} & \it{2.877}
& \it{1.85} & \it{0.4} & \it{86.30} &  \it{2.71} & \it{4.19} &\it{0.25}\\
 M19 & 1.22 & 0.0168 & 0.439 & 6610 & 3.491 & 0.127 & 1.427 & 2.331 & 1.65 & 0.4 & 86.55 &  9.59 & 17.86 &0.95\\
 M20 & 1.22 & 0.0168 & 0.423 & 6632 & 3.565 & 0.127 & 1.431 & 2.451 & 1.75 & 0.4 & 86.37 &   6.77 & 11.83 &0.36\\
 M21 & 1.22 & 0.0168 & 0.407 & 6652 & 3.622 & 0.128 & 1.434 & 2.554 & 1.85 & 0.4 & 86.25 &   3.56 & 5.73 &0.23\\
\hline
 M22 & 1.14 & 0.0112 & 0.412 & 6670 & 3.475 &  0.140 & 1.397 &  3.215 & 1.75 & 0.6 &  86.47 &   8.37 &17.84 & 0.89\\
 M23 & 1.14 & 0.0112 & 0.400 & 6692 & 3.532 & 0.138 & 1.399 & 3.304 & 1.85 & 0.6 & 86.46 &   9.06 & 11.67 &0.67\\
 M24 & 1.18 & 0.0140 & 0.440 & 6657 & 3.548 & 0.145 & 1.414 & 2.882 & 1.75 & 0.6 & 86.45 &   7.23 & 16.29 &0.52\\
 M25 & 1.18 & 0.0140 & 0.427 & 6679 & 3.589 & 0.145 & 1.417 & 2.979 & 1.85 & 0.6 & 86.35 &   4.36 & 9.88 &0.32\\
 \it{M26} & \it{1.20} & \it{0.0168} & \it{0.430} & \it{6603} & \it{3.491} & \it{0.1454} & \it{1.431} & \it{2.972}
& \it{1.85} & \it{0.6} & \it{86.45} &  \it{1.99} & \it{3.79} &\it{0.74}\\
 \it{M27}& \it{1.20} & \it{0.0168} & \it{0.423} & \it{6613} & \it{3.526} & \it{0.1449}
 & \it{1.433} & \it{3.028}& \it{1.90} & \it{0.6} & \it{86.17} & \it{2.33} & \it{3.61} & \it{0.19}\\
 M28 & 1.22 & 0.0168 & 0.458 & 6637 & 3.631 & 0.150 & 1.445 & 2.609 & 1.75 & 0.6 & 86.39 &   7.66 & 10.09 &0.53\\
\noalign{\smallskip}\hline\hline

\end{tabular}
\end{center}
\end{table*}

\begin{table*}
\begin{center}
\renewcommand\arraystretch{1.0}
\caption[]{Same as Table \ref{tmod1}, but for models with diffusion.}
\label{tmod2}
\begin{tabular}{p{0.40cm}cccccccccccccc}
  \hline\hline\noalign{\smallskip}
   Model & $M$ & $(Z/X)_{s}$ &$X_{c}$ & \teff{}& $ L$ & $r_{cc}$& $R$& age&$\alpha$
   & \dov{} &$\Delta\nu$& $\chi_{\nu}^{2}$ & $\chi^{2}_{\nu_\mathrm{corr}}$  & $\chi_{c}^{2} $      \\
   &  (\dsm{})         & &  &      (K)  &  (\dsl{})  & (\dsr{}) & (\dsr{}) & (Gyr) & & &($\mu$Hz)& & & \\
\hline\noalign{\smallskip}
 M29 & 1.26 & 0.0062 & 0.405 & 6591 & 3.516 & 0.087 & 1.440 & 1.484 & 1.75 & 0.0 & 86.45 &  8.22 & 14.74 &0.71\\
 \it{M30} & \it{1.26} & \it{0.0076} & \it{0.383} & \it{6610} & \it{3.573} & \it{0.088} & \it{1.443} & \it{1.596}
& \it{1.85} & \it{0.0} & \it{86.33} &  \it{5.45} & \textit{8.31} &\it{0.38}\\
 M31 & 1.28 &  0.0087 & 0.417 & 6566 & 3.516 & 0.091 & 1.450 & 1.433 & 1.75 & 0.0 & 86.39 &   6.68 & 11.35 &0.73\\
\hline
 M32 & 1.26 & 0.0062 & 0.465 & 6597 & 3.524 & 0.113 & 1.439 & 1.494 & 1.75 & 0.2 & 86.45 &  8.64 & 16.34 &0.67\\
 M33 & 1.26 & 0.0074 & 0.446 & 6618 & 3.589 & 0.113 & 1.443 & 1.603 & 1.85 & 0.2 & 86.32 &   5.87 & 9.52 &0.34\\
 M34 & 1.28 & 0.0085 & 0.475 & 6573 & 3.524 & 0.117 & 1.450 & 1.440 & 1.75 & 0.2 & 86.41 &  6.87 & 12.40 &0.73\\
 \it{M35} & \it{1.28} & \it{0.0099} & \it{0.454} & \it{6594} & \it{3.589} & \it{0.117} & \it{1.416} & \it{1.554}
& \it{1.85} & \it{0.2} & \it{86.29} &  \it{4.30} &\textit{6.79} &\it{ 0.26}\\
 M36 & 1.30 & 0.0092 & 0.505 & 6536 & 3.475 & 0.120 & 1.454 & 1.246 & 1.65 & 0.2 & 86.48 &  9.00 & 17.03 &1.23\\
 M37 & 1.30 & 0.0108 & 0.483 & 6558 & 3.540 & 0.119 & 1.459 & 1.372 & 1.75 & 0.2 & 86.36 &   5.83 & 10.13 &0.74\\
\hline
 M38 & 1.26 & 0.0061 & 0.509 & 6603 & 3.532 & 0.141 & 1.439 & 1.536 & 1.75 & 0.4 & 86.44 &  9.46 & 18.39 &0.62\\
 M39 & 1.26 & 0.0073 & 0.494 & 6626 & 3.597 & 0.140 & 1.441 & 1.639 & 1.85 & 0.4 & 86.41 &  7.58 & 11.26 &0.49\\
 \it{M40} & \it{1.28} & \it{0.0097} & \it{0.499} & \it{6602} & \it{3.606} & \it{0.143} & \it{1.453} & \it{1.595}
& \it{1.85} & \it{0.4} & \it{86.31} &  \it{4.70} &\it{ 8.53} &\it{0.45}\\
 M41 & 1.30 & 0.0092 & 0.542 &  6546 & 3.475 & 0.145 & 1.454 & 1.277 & 1.65 & 0.4 & 86.53 &  9.66 & 17.88 &1.15\\
 M42 & 1.30 & 0.0106 & 0.521 &  6564 & 3.548 & 0.146 & 1.459 & 1.410 & 1.75 & 0.4 & 86.42 &   7.08 & 13.31 &0.69\\
\hline
 M43 & 1.26 & 0.0059 & 0.539 & 6608 & 3.540 & 0.160 & 1.452 & 1.580 & 1.75 & 0.6 & 86.54 &  11.09 &20.15  &0.93\\
 M44 & 1.26 & 0.0071 & 0.519 & 6632 & 3.614 & 0.159 & 1.442 & 1.696 & 1.85 & 0.6 & 86.35 &  7.28 & 13.46 &0.40\\
 M45 & 1.28 & 0.0069 & 0.550 & 6557 & 3.467 & 0.162 & 1.445 & 1.402 & 1.65 & 0.6 & 86.54 &  10.70 & 23.07 &1.26\\
 M46 & 1.28 & 0.0082 & 0.537 & 6583 & 3.540 & 0.161 & 1.448 & 1.521 & 1.75 & 0.6 & 86.46 &   7.72 & 15.74 &0.75\\
 M47 & 1.28 & 0.0094 & 0.525 & 6608 & 3.614 & 0.164 & 1.452 & 1.646 & 1.85 & 0.6 & 86.31 &  5.28 & 10.13 &0.46\\
 \it{M48} & \it{1.28} & \it{0.0135} & \it{0.507} & \it{6537} & \it{3.491} & \it{0.160} & \it{1.458} & \it{1.837}
& \it{1.85} & \it{0.6} & \it{86.25} &  \it{2.19} &\it{3.16}  &\it{0.81}\\
 M49 & 1.29 & 0.0132 & 0.511 & 6577 &  3.595 & 0.162 & 1.462 & 1.763&  1.90 & 0.6 & 86.19 &  3.05 & 4.98 & 0.46\\
 M50 & 1.30 & 0.0104 & 0.542 & 6568 &  3.556 & 0.165 & 1.459 & 1.454 & 1.75 & 0.6 & 86.38 &  6.66 & 13.08 &0.71\\
\hline
 M51 & 1.28 & 0.0133 & 0.519 & 6541 & 3.499 & 0.171 & 1.458 & 1.867 & 1.85 & 0.7 & 86.29 &  4.99 &5.29  &0.80\\
 \it{M52} & \it{1.29} & \it{0.0125} & \it{0.529} & \it{6569} & \it{3.564} & \it{0.170} & \it{1.460} & \it{1.725}
& \it{1.85} & \it{0.7} & \it{86.32} &  \it{3.04} & \it{4.23} &\it{0.61}\\
 M53 & 1.28 & 0.0131 & 0.529 & 6545 & 3.508 & 0.180 & 1.458 & 1.902 & 1.85 & 0.8 & 86.32 &  4.78 & 7.15 &0.79\\
 M54 & 1.29 & 0.0123 & 0.538 & 6573 & 3.573 & 0.181 & 1.459 & 1.758 & 1.85 & 0.8 & 86.32 & 4.26 & 7.21 &0.58\\
\noalign{\smallskip}\hline\hline
\end{tabular}
\end{center}
\end{table*}
\end{document}